\documentclass[aps,twocolumn,,showpacs,superscriptaddress]{revtex4}

\usepackage{epsf}

\begin{document}

\title{Vacuum-field Rabi oscillations in atomically doped carbon nanotubes}
\author{I. V. Bondarev}
\altaffiliation{On leave from the Institute for Nuclear Problems
at the Belarusian State University, Bobruiskaya Str.11, 220050
Minsk, BELARUS} \affiliation{Facult\'{e}s Universitaires
Notre-Dame de la Paix, 61 rue de Bruxelles, 5000 Namur, BELGIUM}
\author{G. Ya. Slepyan}
\author{S. A. Maksimenko}
\affiliation{The Institute for Nuclear Problems, The Belarusian
State University, Bobruiskaya Str.11, 220050 Minsk, BELARUS}
\author{Ph. Lambin}
\affiliation{Facult\'{e}s Universitaires Notre-Dame de la Paix, 61
rue de Bruxelles, 5000 Namur, BELGIUM}

\begin{abstract}
We report a strictly non-exponential spontaneous decay dynamics of
an excited two-level atom placed inside or at different distances
outside a carbon nanotube.~This is the result of strong
non-Markovian memory effects arising from the rapid frequency
variation of the photonic density of states near the nanotube.~The
system exhibits vacuum-field Rabi oscillations when the atom is
close enough to the nanotube surface and the atomic transition
frequency is in the vicinity of the resonance of the photonic
density of states.
\end{abstract}
\pacs{61.46.+w, 73.22.-f, 73.63.Fg, 78.67.Ch}

\maketitle

It has long been recognized that the spontaneous emission rate of
an excited atom is not an immutable property, but that it can be
modified by the atomic environment.~Generally called the Purcell
effect~\cite{Purcell}, the phenomenon is qualitatively explained
by the fact that the local environment modifies the strength and
distribution of the \emph{vacuum} electromagnetic modes with which
the atom can interact, resulting indirectly in the alteration of
atomic spontaneous emission properties.

The Purcell effect took on special significance recently in view
of rapid progress in physics of nanostructures. Here, the control
of spontaneous emission has been predicted to have a lot of useful
applications, ranging from the improvement of existing devices
(lasers, light emitting diodes) to such nontrivial functions as
the emission of nonclassical states of light~\cite{Weisbuch}.~In
particular, the enhancement of the spontaneous emission rate can
be the first step towards the realization of a thresholdless
laser~\cite{Pelton} or a single photon source~\cite{Gerard}. The
possibility to control atomic spontaneous emission was shown
theoretically for microcavities and
microspheres~\cite{Dung,Kimble}, optical fibers~\cite{Klimov},
photonic crystals~\cite{John}, semiconductor quantum
dots~\cite{Sugawara}. Recent technological progress in the
fabrication of low-dimensional nanostructures has enabled the
experimental investigation of spontaneous emission for
microcavities~\cite{Schniepp}, photonic crystals~\cite{Petrov},
quantum dots~\cite{Gayral}.

Recent successful experiments on encapsulation of single atoms
into single-wall carbon nanotubes (CNs)~\cite{Jeong} and their
intercalation into single-wall CN bundles~\cite{Shimoda} stimulate
an analysis of atomic spontaneous emission properties of such
systems.~Typically, there may be two qualitatively different
regimes of the interaction of the excited atomic state with a
\emph{vacuum} electromagnetic field in the vicinity of the
CN.~They are the weak coupling regime and the strong coupling
regime.~The former is characterized by the monotonous exponential
decay dynamics of the upper atomic state with the decay rate
altered compared with the free-space value.~The latter~is, in
contrast, strictly non-exponential and~is characterized by
reversible Rabi oscillations where the energy of the initially
excited atom is periodically exchanged between the atom and the
field.~In this Letter, we develop the quantum theory of the
spontaneous decay of a two-level atom near a CN, derive the
evolution equation of the upper state of the system and, by
solving~it numerically, demonstrate the strictly
\emph{non-exponential} spontaneous decay dynamics in the case
where the atom is close enough to the CN surface.~In certain
cases, the system exhibits \emph{vacuum-field Rabi oscillations}
-- a result already detected for quasi-2D excitonic and
intersubband electronic transitions in semiconductor quantum
microcavities~\cite{Weisbuch1,Sorba} and never reported for
atomically doped CNs so far.

The quantum theory of the spontaneous decay of excited atomic
states in the presence of a CN involves an electromagnetic field
quantization procedure.~Such~a procedure faces difficulties
similar to those in quantum optics of 3D Kramers-Kronig dielectric
media where the canonical quantization scheme commonly used does
not work since, because of absorption, operator Maxwell equations
become non-Hermitian~\cite{Vogel}.~As a result, their solutions
cannot be expanded in power orthogonal modes and the concept of
modes itself becomes more subtle.~We, therefore, use an
alternative quantization scheme developed in Ref.~\cite{Dung},
where Fourier-images of electric and magnetic fields are
considered as quantum mechanical observables of corresponding
electric and magnetic field operators.~The latter ones satisfy the
Fourier-domain operator Maxwell equations modified by the presence
of a~so-called operator noise current written in terms of a vector
bosonic field operator and an imaginary dielectric
permittivity.~This operator is responsible for correct commutation
relations of the electric and magnetic field operators in the
presence of medium-induced absorbtion.~The electric and magnetic
field operators are then expressed in terms of a continuum set~of
the bosonic fields by means of the convolution of the operator
noise current with a \emph{classical} electromagnetic field Green
tensor of the system.~The bosonic field operators create and
annihilate single-quantum electromagnetic medium excitations.~They
are defined by their commutation relations and play the role of
the fundamental dynamical variables in terms of which the
Hamiltonian of the composed system "electromagnetic field +
dissipative medium" is written in a standard secondly quantized
form.

We consider a two-level atom positioned at the point
$\mathbf{r}_{A}$ near an infinitely long single-wall~CN. The atom
interacts with a quantum electromagnetic field via an electric
dipole transition of frequency $\omega_{A}$.~The atomic dipole
moment, $\mathbf{d}\!=\!d_{z}\textbf{e}_{z}$, is assumed to be
directed along the CN axis assigned by the unit vector
$\textbf{e}_{z}$ of the orthonormal cylindric basis
$\{\mathbf{e}_{r},\mathbf{e}_{\varphi},\mathbf{e}_{z}\}$.~The
contribution of the transverse dipole moment orientation is
suppressed because of the strong depolarization of the transverse
field in an isolated CN~\cite{Li}.~The Hamiltonian of the system
is given by
\[
\hat{H}=\int\!d\mathbf{R}\!\int_{0}^{\infty}\!\!\!d\omega\,
\hbar\omega\,\hat{f}^{\dag}(\mathbf{R},\omega)\hat{f}(\mathbf{R},\omega)
+{1\over{2}}\,\hbar\omega_{A}\,\hat{\sigma}_{z}
\]\vspace{-0.5cm}
\begin{equation}
-[\,\hat{\sigma}^{\dag}\,\hat{E}^{(+)}_{z}(\mathbf{r}_{A})\,d_{z}+\mbox{h.c.}\,]
\label{Ham}
\end{equation}
with the three terms representing the electromagnetic field
(modified by the presence of the CN), the two-level atom and their
interaction (within the framework of electric dipole and rotating
wave approximations~\cite{Dung}), respectively.~The operators
$\hat{f}^{\dag}(\mathbf{R},\omega)$ and
$\hat{f}(\mathbf{R},\omega)$ are the scalar bosonic field
operators defined on the CN surface
($\mathbf{R}=\!\{R_{cn},\phi,Z\}$~is the radius-vector of an
arbitrary point of the CN surface).~They play the role of the
fundamental dynamical variables of the field subsystem and satisfy
the standard bosonic commutation relations.~The Pauli operators
$\hat{\sigma}_{z}$, $\hat{\sigma}$, $\hat{\sigma}^{\dag}$ describe
the atomic subsystem. The operators
$\hat{\mathbf{E}}^{(\pm)}(\mathbf{r}_{A})$ represent the electric
field the atom interacts with.~For an arbitrary
$\mathbf{r}\!=\!\{r,\varphi,z\}$, they are defined as
$\hat{\mathbf{E}}(\mathbf{r})=\hat{\mathbf{E}}^{(+)}(\mathbf{r})+
\hat{\mathbf{E}}^{(-)}(\mathbf{r})$ with
\begin{equation}
\hat{\mathbf{E}}^{(+)}(\mathbf{r})=\int_{0}^{\infty}
\underline{\hat{\mathbf{E}}}(\mathbf{r},\omega)\;d\omega\,,\;\;\;
\hat{\mathbf{E}}^{(-)}(\mathbf{r})=[\hat{\mathbf{E}}^{(+)}(\mathbf{r})]^{\dag}.
\label{Ew}
\end{equation}
Here, $\underline{\hat{\mathbf{E}}}$ satisfies the Fourier-domain
Maxwell equations
\[
\nabla\times\underline{\hat{\mathbf{E}}}(\mathbf{r},\omega)=
ik\,\underline{\hat{\mathbf{H}}}(\mathbf{r},\omega),
\]\vspace{-0.5cm}
\begin{equation}
\nabla\times\underline{\hat{\mathbf{H}}}(\mathbf{r},\omega)=
-ik\,\underline{\hat{\mathbf{E}}}(\mathbf{r},\omega)+
{4\pi\over{c}}\underline{\hat{\mathbf{I}}}(\mathbf{r},\omega),\label{Maxwell}
\end{equation}
where $\underline{\hat{\mathbf{H}}}$ stands for the magnetic field
operator [defined by analogy with Eq.~(\ref{Ew})], $k=\omega/c$,
and
\begin{equation}
\underline{\hat{\mathbf{I}}}(\mathbf{r},\omega)\!=\!\!\!\int\!\!d\mathbf{R}\,
\delta(\mathbf{r}-\mathbf{R})\,\underline{\hat{\mathbf{J}}\!}\,(\mathbf{R},\omega)\!
=\!2\underline{\hat{\mathbf{J}}\!}\,(R_{cn},\varphi,z,\omega)\delta(r-R_{cn})
\label{Irw}
\end{equation}
with $\underline{\hat{\mathbf{J}}\!}=
\!\sqrt{\hbar\omega\mbox{Re}\sigma_{zz}(\mathbf{R},\omega)/\pi}
\hat{f}(\mathbf{R},\omega)\textbf{e}_{z}$ being the operator noise
current density associated with CN-induced absorbtion
($\sigma_{zz}$ is the CN surface axial conductivity per unit
length; in describing CN electronic properties we use the model of
the axial conductivity~\cite{Slepyan}).

From Eq.~(\ref{Maxwell}) it follows that
\begin{equation}
\underline{\hat{\mathbf{E}}}(\mathbf{r},\omega)=
i{4\pi\over{c}}\,k\!\int\!\!d\mathbf{R}\,\mathbf{G}(\mathbf{r},\mathbf{R},\omega)
\!\cdot\!\underline{\hat{\mathbf{J}}\!}\,(\mathbf{R},\omega)
\label{Erw}
\end{equation}
[and $\underline{\hat{\mathbf{H}}}=(ik)^{-1}
\nabla\times\underline{\hat{\mathbf{E}}}$ accordingly], where
$\mathbf{G}(\mathbf{r},\mathbf{R},\omega)$ is the Green tensor of
the \emph{classical} electromagnetic field in the vicinity of the
CN. Its components satisfy the equation
\begin{equation}
\sum_{\alpha=r,\varphi,z}\!\!\!
\left(\mathbf{\nabla}\!\times\mathbf{\nabla}\!\times-\,k^{2}\right)_{\!z\alpha}
G_{\alpha z}(\mathbf{r},\mathbf{R},\omega)=
\delta(\mathbf{r}-\mathbf{R}), \label{GreenequCN}
\end{equation}
together with the radiation conditions at infinity and~the
boundary conditions on the CN surface.~The latter ones are
obtained from the classical electromagnetic field bou\-ndary
conditions derived in Ref.~\cite{Slepyan} by means of simple
relations valid for $\mathbf{r}\!\ne\!\mathbf{r}_{A}$:
$E_{\alpha}(\mathbf{r},\omega)\!=\!ikG_{\alpha
z}(\mathbf{r},\mathbf{r}_{A},\omega)$ and
$H_{\alpha}\!=\!(ik)^{-1}\!\sum_{\beta,\gamma}
\epsilon_{\alpha\beta\gamma}\nabla_{\beta}E_{\gamma}$
($\epsilon_{\alpha\beta\gamma}$ is the totally antisymmetric unit
tensor).~The Hamiltonian~(\ref{Ham}) along with
Eqs.~(\ref{Ew})--(\ref{GreenequCN}) is the modification of the
Jaynes--Cummings model~\cite{Vogel} for an atom in the vicinity of
a~solitary~CN.

When the atom is initially in the upper state and the field
subsystem is in vacuum, the time-dependent wave function of the
whole system can be written as
\[
|\psi(t)\rangle=C_{u}(t)\,e^{-i(\omega_{A}/2)t}|u\rangle|\{0\}\rangle
\]\vspace{-0.5cm}
\begin{equation}
+\int\!d\mathbf{r}\!\int_{0}^{\infty}\!\!\!d\omega\,
C_{l}(\mathbf{r},\omega,t)\,e^{-i(\omega-\omega_{A}/2)t}
|l\rangle|\{1(\mathbf{r},\omega)\}\rangle, \label{wfunc}
\end{equation}
where $|u\rangle$ and $|l\rangle$ are the upper and lower atomic
states, respectively, $|\{0\}\rangle$ is the vacuum state of the
field subsystem, $|\{1(\mathbf{r},\omega)\}\rangle$ is its excited
state where the field is in a single-quantum Fock state.~In view
of Eqs.~(\ref{Ew}) and~(\ref{Erw}), the Schr\"{o}dinger equation
with Hamiltonian~(\ref{Ham})~and wave function~(\ref{wfunc})
yields the following evolution law for the oc\-cupation
probability amplitude $C_{u}$ of the upper state
\begin{equation}
C_{u}(\tau)=1+\int_{0}^{\tau}\!K(\tau-\tau^{\prime})\,
C_{u}(\tau^{\prime})\,d\tau^{\prime}\,, \label{Volterra}
\end{equation}
\begin{equation}
K(\tau-\tau^{\prime})={\hbar\Gamma_{0}(x_{A})\over{4\pi
x_{A}^{3}\gamma_{0}}}\!\int_{0}^{\infty}\!\!\!\!\!dx\,x^{3}\xi(x)
{e^{-i(x-x_{A})(\tau-\tau^{\prime})}-1\over{i\,(x-x_{A})}}.
\label{kernel}
\end{equation}
Here, $x\!=\!{\hbar\omega/2\gamma_{0}}$ and
$\tau\!=\!2\gamma_{0}t/\hbar$ are the dimensionless frequency and
time, respectively, with $\gamma_{0}\!=\!2.7$~eV being the carbon
nearest neighbor hopping integral appearing in the CN surface
axial conductivity in the noise current in Eq.~(\ref{Irw}),
$\xi(x)\!=\!\Gamma(x)/\Gamma_{0}(x)$ is the relative density (with
respect to vacuum) of photonic states near the CN with
$\Gamma(x)\!=\!(8\pi d_{z}^{2}/\hbar
c^{2})(2\gamma_{0}x/\hbar)^{2}\mbox{Im}\,G_{zz}(\mathbf{r}_{A},\mathbf{r}_{A},x)$
being the rate of the \emph{exponential} spontaneous decay of the
atom near the CN and $\Gamma_{0}(x)$, its vacuum counterpart,
given by the same formula with
$\mbox{Im}\,G_{zz}^{0}(x)\!=\!(1/6\pi c)(2\gamma_{0}x/\hbar)$.~The
function $\xi(x)$ was first derived and analyzed in
Ref.~\cite{Bondarev02}. Eq.~(\ref{Volterra}) is a well-known
Volterra integral equation of the second kind. In our case, it
describes the spontaneous decay dynamics of the excited two-level
atom in the vicinity of the CN. All the CN parameters that are
relevant for the spontaneous decay are contained in the classical
electromagnetic field Green tensor in Eq.~(\ref{kernel}).

In the case where the Markovian approximation is applicable, or,
in other words, when the atom-field coupling strength is weak
enough for atomic motion memory effects to be insignificant, so
that they may be neglected, the factor
$\{\exp[-i(x-x_{A})(\tau-\tau{^\prime})]-\!1\}/i(x-x_{A})$ in
Eq.~(\ref{kernel}) may be replaced by its long-time approximation
$-\pi\delta(x-x_{A})+i{\cal{P}}/(x-x_{A})$ (${\cal{P}}$ denotes a
principal value).~Then,
$K(\tau-\tau{^\prime})\!=\!-\hbar\Gamma(x_{A})/4\gamma_{0}+i\Delta(x_{A})$
with $\Delta(x_{A})\!=\![\hbar\Gamma_{0}(x_{A})/4\pi
x_{A}^{3}\gamma_{0}]\,{\cal{P}}\!
\int_{0}^{\infty}\!dx\,x^{3}\xi(x)/(x-x_{A})$, and
Eq.~(\ref{Volterra}) yields
$C_{u}(\tau)\!=\!\exp[(-\hbar\Gamma/4\gamma_{0}+i\Delta)\tau]$ ---
the exponential decay dynamics of the [shifted by $\Delta(x_{A})$]
upper atomic level with the rate $\Gamma(x_{A})$. This case was
analyzed in Ref.~\cite{Bondarev02}.

\begin{figure}[t]
\epsfxsize=8.65cm \centering{\epsfbox{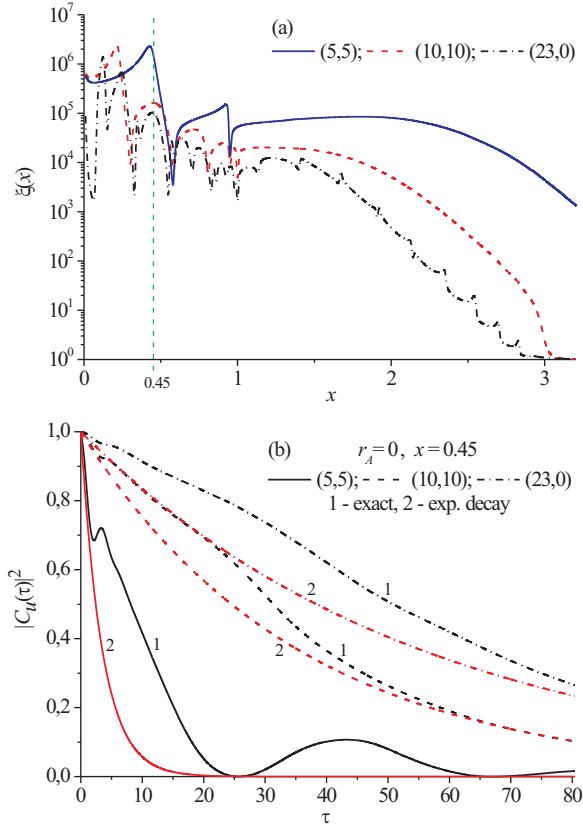}}\vskip-0.5cm
\caption{(Color online) Relative density of photonic states (a)
and upper-level spontaneous decay dynamics (b) for the atom in the
center of different CNs.} \label{fig1}
\end{figure}

\begin{figure}[t]
\epsfxsize=8.65cm \centering{\epsfbox{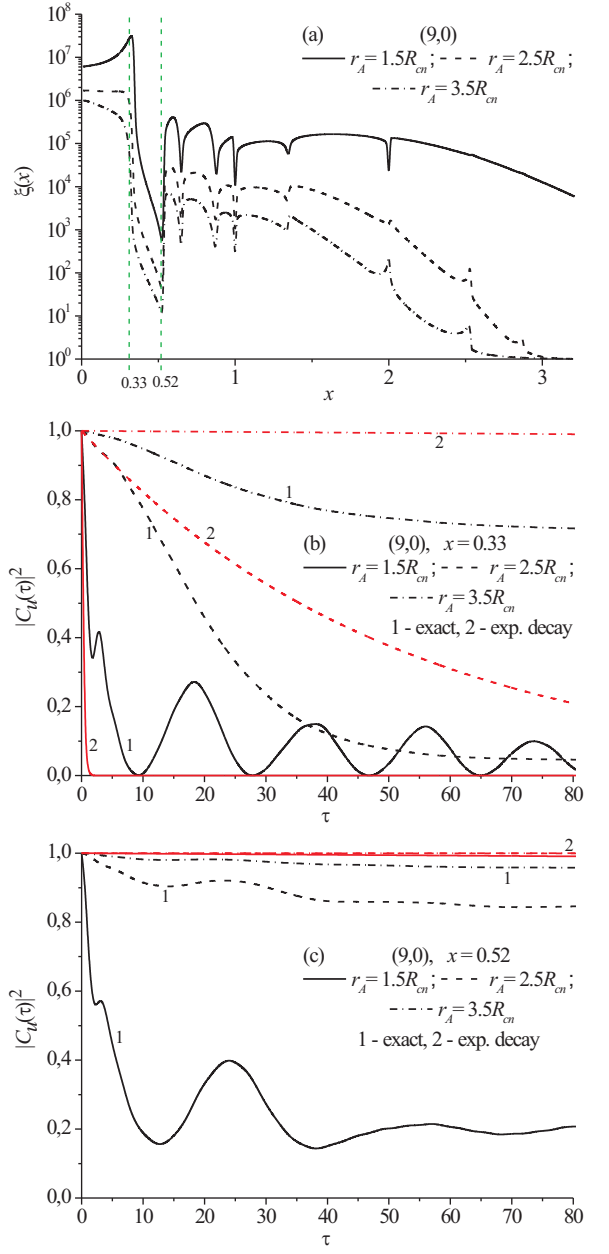}}\vskip-0.5cm
\caption{(Color online) Relative density of photonic states (a)
and upper-level spontaneous decay dynamics (b, c) for the atom at
different distances outside (9,0) CN.} \label{fig2}
\end{figure}

To take non-Markovian effects into account, we have solved
Eqs.~(\ref{Volterra}) and~(\ref{kernel}) numerically. The
\emph{exact} time evolution of the occupation probability
$|C_{u}(\tau)|^{2}$ of the upper state was obtained for the atom
placed in the center inside and at different distances outside
achiral CNs of different radii.~The factor $\xi(x)$ in
Eq.~(\ref{kernel}) was computed in the same manner as it was done
in Ref.~\cite{Bondarev02}. The free-space spontaneous decay rate
was approximated as
$\Gamma_{0}(x_{A})\!\approx\!\alpha^{3}2\gamma_{0}x_{A}/\hbar$
with $\alpha\!=\!1/137$ being the fine-structure
constant~\cite{Davydov}.

Figure~\ref{fig1}(a) presents $\xi(x)$ for the atom in the center
of the (5,5), (10,10) and (23,0) CNs.~It is seen to decrease with
increasing the CN radius, representing the decrease of the
atom-field coupling strength as the atom moves away from the CN
surface~\cite{Bondarev02}. To calculate $|C_{u}(\tau)|^{2}$ in
this particular case, we have fixed $x_{A}\!=\!0.45$ (indicated by
the vertical dashed line), firstly, because this transition is
located within the visible light range $0.305\!<\!x\!<\!0.574$,
secondly, because this is the approximate peak position of
$\xi(x)$ for all the three CNs.~The functions $|C_{u}(\tau)|^{2}$
calculated are shown in Figure~\ref{fig1}(b) in comparison with
those obtained in the Markovian approximation yielding the
exponential decay.~The actual spontaneous decay dynamics is
clearly seen to be non-exponential.~For the small radius (5,5) CN,
Rabi oscillations are observed, indicating a strong atom-field
coupling regime related to strong non-Markovian memory effects.
With increasing the CN radius, the decay dynamics approaches the
exponential one with the decay rate enhanced by several orders of
magnitude compared with that in free space.

Figure~\ref{fig2}(a) shows $\xi(x)$ for the atom outside the (9,0)
CN at the different distances from its surface.~The vertical
dashed lines indicate the atomic transitions for which the
functions $|C_{u}(\tau)|^{2}$ in Figures~\ref{fig2}(b)
and~\ref{fig2}(c) were calculated.~Both transitions belong to the
visible light range, however $x_{A}\!=\!0.33$ is the position of a
peak (for the shortest atom-surface distance) while
$x_{A}\!=\!0.52$ is the position of a dip of $\xi(x)$.~Very clear
underdamped Rabi oscillations are seen for the shortest
atom-surface distance at $x_{A}\!=\!0.33$ [Figure~\ref{fig2}(b)],
indicating strong atom-field coupling with strong
non-Markovity.~For $x_{A}\!=\!0.52$ [Figure~\ref{fig2}(c)], though
$\xi(0.52)$ is comparatively small, the spontaneous decay dynamics
is still non-exponential, approaching the exponential one only
when the atom is far enough from the CN surface. Similar to what
takes place in photonic crystals~\cite{John}, this is because of
the fact that, due to the rapid variation of $\xi(x)$ in the
neighborhood of this frequency, the correlation time of the
electromagnetic vacuum is not negligible on the time scale of the
evolution of the atomic system, so that atomic motion memory
effects are important and the Markovian approximation in
Eq.~(\ref{kernel}) is inapplicable.

The effects predicted will yield an additional structure in
optical absorbance/reflectance spectra (see, e.g.,~\cite{Li}) of
atomically doped CNs in the vicinity of the energy of an atomic
transition. Weak non-Markovity of the spontaneous decay
(non-exponential decay with no Rabi oscillations) will cause an
\emph{asymmetry} of an optical spectral line-shape (similar to
exciton optical absorbtion line-shape in quantum
dots~\cite{Bondarev03}).~Strong non-Markovity of the spontaneous
decay (non-exponential decay with fast Rabi oscillations)
originates from strong atom-vacuum-field coupling with the upper
state of the system splitted into two "dressed" states.~This will
yield a \emph{two-component} structure of optical spectra similar
to that observed for excitonic and intersubband electronic
transitions in semiconductor quantum
microcavities~\cite{Weisbuch1,Sorba}.

To conclude, we predict the existence of frequency ranges where,
similar to semiconductor microcavities~\cite{Sorba} and photonic
band-gap materials~\cite{John}, CNs qualitatively change the
character of atom-electromagnetic-field interaction, yielding
strong atom-field coupling --- an important phenomenon necessary,
e.g., for quantum information processing~\cite{Abstreiter}.~The
present Letter dealt with the simplest manifestation of this
general phenomenon --- vacuum-field Rabi oscillations in the
atomic spontaneous decay dynamics near a single-wall carbon
nanotube.~Such a strictly non-exponential decay dynamics gives
place to the exponential one if the atom moves away from the CN
surface.~Thus, the atom-field coupling strength and the character
of the spontaneous decay dynamics, respectively, may be controlled
by changing the distance between the atom and the CN surface by
means of a proper preparation of atomically doped CNs.~This opens
routes for new challenging nanophotonics applications of
atomically doped CN systems as various sources of coherent light
emitted by dopant atoms.~We emphasize that similar manifestations
of strong atom-field coupling may occur in many other
atom-electromagnetic-field interaction processes in the presence
of CNs, such as, e.g., dipole-dipole interaction between atoms by
means of a vacuum photon exchange~\cite{Agarwal}, or cascade
spontaneous transitions in three-level atomic
systems~\cite{Dalton}.

Discussions with Prof. I.~D.~Feranchuk are gratefully
acknowledged. I.B.~thanks the Belgian OSTC. The work was performed
within the framework of the Belgian PAI-P5/01 project.

\end{document}